# The Influence of Trust and subjective Norms on Citizens' Intentions to Engage in E-participation on E-government Websites

**Abdullah Alharbi**
Faculty of Engineering and Information Technology
University of Technology Sydney, Australia
Email: Abdullah.m.alharbi@student.uts.edu.au

**Dr. Kyeong Kang**
Faculty of Engineering and Information Technology
University of Technology Sydney, Australia
Email: kyeong.kang@uts.edu.au

**Prof. Igor Hawryszkiewycz**
Faculty of Engineering and Information Technology
University of Technology Sydney, Australia
Email: Igor.Hawryszkiewycz@uts.edu.au

**ABSTRACT**

Advancements in web technology have revolutionised the way citizens interact with governments. Unlike traditional methods of communication between citizens and governments, e-participation via e-government websites enhances communication and enables citizens to become actively involved in the policy-making process. Despite the growing importance of e-participation, the potential factors influencing citizens' engagement in e-participation have not yet been investigated. Using responses obtained from Saudi citizens, this study examines a number of factors that may influence the intentions of citizens to engage in e-participation activities on e-government websites. The results suggest that the factors of trust and subjective norms have a significant impact on citizens' intentions to engage in e-participation activities.

## 1 INTRODUCTION

The growing use of web technologies has significantly influenced how governments work and provide services to citizens. It has also affected the manner in which citizens interact and communicate with their governments. Government 2.0 utilises Web 2.0 technologies to socialise government online services, processes and data (Nam 2011). Collaborative technologies permit two-way interactions between citizens and governments through online comments, live chats and message threads (Nam 2012). The increased use of web technologies has also influenced how governments perform their functions. The wide range of available social networking sites and the tools used by e-government websites, including forums and blogs, also provide insights into citizens' perceptions and opinions. E-participation services offer a form of two-way communication between citizens and governments and trust and subjective norms are key to making these effective.

In recent years, there have been significant advancements in the area of e-participation. Previous studies have explored the process of e-participation and the models used to enhance it (Macintosh and Whyte 2008; Hu et al. 2014). However, to date, little attention has been given to the factors affecting citizens' engagement in e-participation and large gaps therefore exist in relation to our understanding of this area (Reddick 2011). A continuing challenge for governments using online tools is the low level of citizen participation because the full benefits of e-participation cannot be realised until citizens actually engage with and use these tools. A number of e-participation studies have highlighted citizens' lack of participation in e-government services (Nam 2012; Reddick 2011). Reddick (2011) asserts that citizens are highly unlikely to use the Internet, particularly e-government websites, for participatory and counselling activities. However, there is little information available on what deters citizens from using e-participation services or, on the other hand, what facilitates such participation. Such unsolved issues pose practical challenges for e-government leaders who need to make informed decisions in relation to e-participation tools in order to provide quality services to citizens.

Previous studies have shown that various factors affect the intentions of citizens to use e-participation tools (Hung et al. 2006; Nam 2011; Scherer and Wimmer 2014; Kang and Ng 2015). However, these studies have mainly considered Western countries and little research has been undertaken in Saudi Arabia. The Saudi government places immense importance on this subset of e-government tools, with the aim of acquiring the benefits of e-participation and strengthening the relationship between the government and its citizens. Reflecting the importance of e-government tools to the government, the Saudi's second e-government action plan (2012–2016) made e-participation a strategic work stream (Yesser 2012). A recent report from the United Nations (UN) (2014) revealed that Saudi Arabia ranked 36th in providing e-government services; however, despite significant investments being made in e-government services, Saudi Arabia ranked 51st in relation to e-participation, reflecting a low level of citizen engagement in e-participation activities.

Since 2012, the Saudi government has actively implemented e-participation tools through the 'e-government second plan' project on their websites to engage citizen participation. The government's intention was to address the increasing of citizens' intentions to engage in e-participation will assist local government leaders to better understand the major issues facing e-participation and enable analysis of the extent to which citizens' online engagement reflect community interests.

As e-participation is an emerging, but underexplored, area of research in terms of citizen participation, the aim of this study is to investigate trust and subjective norms as the underlying factors influencing citizens' intentions to engage in e-participation. Specifically, this research attempts to answer the following question: how do the key factors (trust and subjective norms) affect citizens' intentions to engage in e-participation on e-government websites in Saudi Arabia? This paper contributes to the knowledge in the area of e-participation by investigating the role of trust and subjective norms in the context of e-participation. The study commences with a literature review and, following this, the theoretical background for the research is considered. A research model is then proposed and a hypothesis developed. Next, the approach adopted in this study is outlined and the findings of the study are set out. Finally, a discussion of the results is undertaken and conclusions are drawn.

## 2  LITERATURE REVIEW

### 2.1 E-Participation

E-Participation enhances communications between governments and citizens and enables citizens to become more involved in the policy-making process. Hung et al. (2006) discovered that citizens engage in e-government websites through e-participation services. Citizens can use tools provided on e-government websites to express their opinions to governments. According to Ali et al. (2015), 'E-Participation is the use of technology to enhance participation between different stakeholders and governments'. There are various levels of e-participation for citizen engagement; for example, the Organization for Economic Cooperation and Development's (OECD's) 2001 model for e-participation incorporates information, consultation and active participation. Similarly, the United Nation's (UN) e-participation 2012 model was based on e-information, e-consultation and decision making. Based on the OECD and UN models, Macintosh (2004) categorised e-participation service as having three levels: e-enabling, e-engaging and e-empowering. Empowering citizens is a particularly important level, as it can provide bottom-up ideas for the e-participation process. E-government websites use different communication tools for e-participation services (Reddick 2011), including email, blogs, social networking sites, surveys, chats and polls. According to Nam (2011), e-participation refers to public involvement in government services through electronic means, such as social media. In the context of the current study, e-participation was defined as the use of information and communication technology (ICT) to enhance interactions between governments and citizens that facilitate public participation in e-government processes, such as consultation, proposing ideas, decision making, policy making, administration, service delivery and easy access to public information.

### 2.2 Trust

E-participation services refer to two-way communications between citizens and government. Thus, trust is essential in ensuring effective e-participation between citizens and governments. Trust refers to the extent to which an individual believes that another individual or group will act in a favourable manner (Mayer et al. 1995). Previous researchers have suggested that two important dimensions drive trust in e-government context; that is, trust in the Internet and trust in the government (McKnight et al. 2002; Carter and Belanger 2005; Belanger and Carter 2008; Schaupp et al. 2010). In addition, analysis of trust from psychology, sociology and social psychology is to be considered when analysing trust in e-participation, such as social trust (Bond 2006; Scherer and Wimmer 2014).

Several studies have argued that trust significantly affects citizens' intentions to use e-government services (Carter and Belanger 2005; AlAwadhi and Morris 2009; Alomari et al. 2012; Al-Sobhi et al. 2011a). Warkentin et al. (2005) pointed out that citizens' trust is a fundamental motivation in the use of e-government services. However, very few empirical studies have examined the role of trust in citizens' intention to engage in e-participation. Trust in governments enables citizens to actively engage in e-participation, as it encourages citizens to have a sense of cooperation with government (Lee and Kim 2014). E-participation tools influences citizens' intention to use the site (Lee and Kin 2014). Trust was conceptualised as a single construct; that is, trust in the government and it was hypothesised that there was a positive relationship between trust in the government and citizens' e-participation use. The previous studies supported that citizens with a high level of trust in the government were more motivated to actively participate in government initiated e-participation tools, as they had a sense of cooperating with the government. Further, citizens who have a higher level of trust in the government were confident that their opinions would be valued and that any time spent engaging in e-participation would not be a wasted effort (Alharbi and Kang 2014); thus, these citizens were more willing to engage in e-participation. In addition, Scherer and Wimmer (2014) showed that trusting the e-participation tools positively affected intended usage. The relationship between trust and citizens' intentions to engage in e-participation activities has been considered to be potential risk and created uncertainties in citizens considering using e-participation (Belanger & Carter 2008). Bélanger and Carter (2005, 2008) found that trust (in the government and in the Internet) had positive effects on individuals' intention to use e-government services, including e-participation services. A study conducted by Reddick (2011) revealed that citizens of the United States (US) were less confident in becoming involved in consultative and participatory activities, as they had less trust for their government. Nam (2014) examined the relationship between trust in government and e-government usage in the US and found that in relation to e-government services, trust in the government was more important than trust in technology. Thus, trust is one of the most important factors influencing citizens' intentions to engage in e-participation activities. However, further research in needed on the impact of trust in the e-participation context.

### 2.3 Subjective Norms

Subjective norms refer to the social influences that impact upon the performance or non performance of behaviours (Ajzen 1991). The introduction of Web 2.0 technologies, including social networking sites, blogs and wiki websites, has allowed for online communication and the exchange of information with the government. Citizens are increasingly more willing to interact with government agencies online (Tapscott et al. 2008). Collaborative technologies permit a two-way interaction between citizens and governments through online comments, live chats and message threads (Nam 2012). Khasawneh and Abu-Shanab (2013) studied the impact of the Jordan e-government page on Facebook and found a strong level of engagement with citizens. McGrath et al. (2011) highlighted the scope of online social networking in e-participation use in Europe, the Middle East and Latin America. Their findings showed that online social networking allowed for the fast mobilisation of citizens and the transfer of immediate information to help managing field situations, such as demonstrations and campaigning to elect political candidates. Al-Hujran and Al-Debei (2014) proposed a model on the basis of the Theory of Planned Behaviour (TPB) and Technology Acceptance Model (TAM) and found that attitudes, subjective norms and perceived behavioural controls positively affected citizens' intentions to engage with the eDemocracy in Jordan. However, subjective norms have been found to have a significant effect on citizen's intention to use eDemocracy tools in Jordan e-government websites. Soon and Soh (2014) highlighted how web 2.0 technologies enhance mutuality, propinquity and empathy in government-citizen communications on a daily basis; for example, government officials are able to assess public sentiments via the comments and feedback of Facebook users in relation to potential disagreements on specific issues and policies.

## 3 THEORITICAL BACKGROUND AND RESEARCH MODEL

The basic idea behind e-participation is to allow citizens to interact with their government through the Internet and to encourage them to express their opinions and participate in decisions. Thus, citizens' trust and subjective norms are significant factors in the intentions of citizens to engage in such services. Previous studies have analysed the role of trust in e-government services (Gefen et al. 2005; Welch et al. 2005; Bélanger and Hiller 2006). Citizen trust is understood to be a combination of beliefs about the attributes of the government agency, the communication channel (e.g., the Internet) and social trust (Blind 2006; Kang and Kovacevic 2012; Scherer and Wimmer 2014). Trust in the government refers to citizens' beliefs that the trustee has characteristics that are beneficial for him or her. Trust in the Internet refers to the extent to which citizens trust the competency and security of the Internet (Bélanger and Carter 2008). Social trust 'refers to citizens' confidence in each other as members of a social community, [and] is inseparable from the notion of political trust' (Blind 2006). In the context of e-participation, the concept of trust is considered as the citizens' expectation that e-participation tools and websites are responsible and reliable. In the current study, trust in the government, trust in the Internet and social trust are all considered to be significant aspects of citizens' trust in relation to the use of e-participation.

A wide range of available social networking sites and tools (e.g., forums and blogs) are being used by e-governments to gain insights into citizens' perceptions and opinions. The key predictor of actual behaviour is behavioural intention as determined by subjective norms towards that behaviour (Fishbein and Ajzen 1975). Subjective norms refer to an individual being socially influenced by the opinions of others (e.g., family members, friends and colleagues) and the fact that these opinions affect performance behaviours (Ajzen 1991). In this study, subjective norms refer to social influence received to engage in e-participation activities. Previous researchers have identified three important determinants for subjective norms: family members, friends/colleagues and the media (Taylor and Todd 1995; Lau 2004; Hung et al. 2006; Lin 2007). Figure 1 shows the research model.

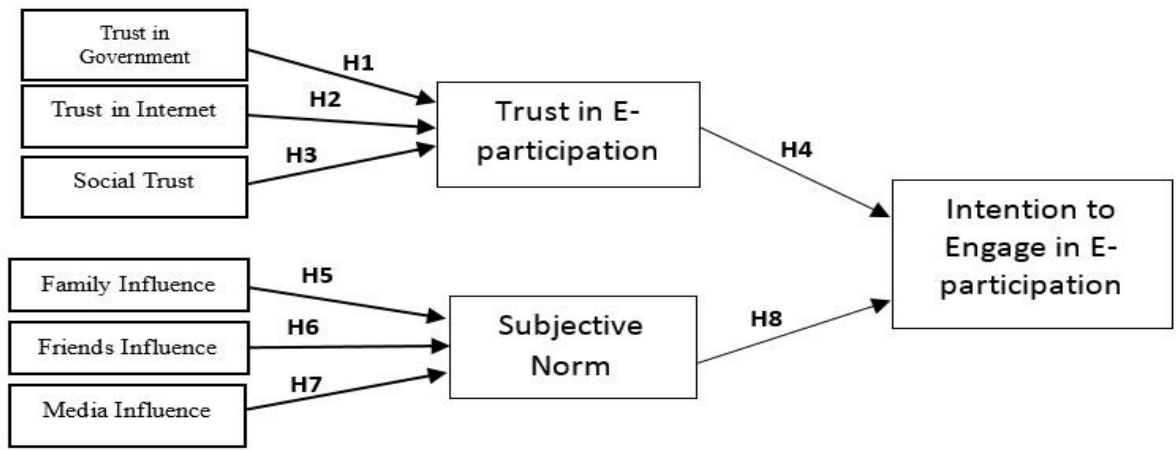

*Figure 1: Research Model*

### 3.1 Hypothesis Development

#### 3.1.1　Trust

Trust between citizens and the government is an important factor in ensuring the successful acceptance of e-government services (Warkentin et al. 2002; Reddick 2011; Lee and Kim 2014). Carter and Bélanger (2005) highlight that 'citizens must have confidence in both the government and the enabling technologies'. Warkentin et al. (2005) point out that trust is a crucial factor in the use of e-government services by citizens. Bélanger and Carter (2008) proposed and tested a model for trust and found that people's disposition to trust, which is parallel to the trust propensity element in the general model described in Colquitt et al. (2007), indeed affected the respondents' level of trust in e-government services. People who have higher dispositions to trust in things in general trust e-government services more than others. Furthermore, Bélanger and Carter (2008) found that trust in the Internet itself and trust in government are separate, important factors that also affected trust in e-government services, such as e-participation services. Later, Alsaghier et al. (2010) investigated the impact of trust in e-government services on citizens' intention to use e-government services. Their findings show a significant positive effect of trust in e-government service on citizens' intention to use these services. Moreover, they found that citizens' trust in e-government services is affected by citizens' position to trust, trust in government and trust in the Internet. In another study, Alzahrani (2011) found that trust in government and the Internet affects Saudi citizens' intention to use e-government services. In the e-participation context, Reddick (2011) found that trust in government affects citizens' attitude towards e-participation services. Previous empirical studies have explored the role of trust towards citizens' intention to engage in e-participation. Lee and Kim (2014), for example, note that trust in government enables citizens to actively engage in e-participation because this trust encourages citizens to have a sense of cooperation with the government. Scherer and Wimmer (2014) claim that social trust in e-participation positively affects the intended usage and citizens are, therefore, more likely to engage in e-participation activities if they trust their governments to take their opinions into consideration. Moreover, trust in the Internet is important because there are many web 2.0 tools used in e-participation activities, such as social networking sites. Trust is a crucial factor in the use of e-participation activities and, as a result, the following hypotheses are developed.

**H1:** Trust in Government positively effects on citizens' trust in e-participation.

**H2:** Trust in the Internet positively effects on citizens' trust in e-participation

**H3:** Social trust positively effects on citizens' trust in e-participation.

**H4:** Citizens' trust in e-participation has a positive relationship with citizen's intention to engage in E-participation.

### 3.1.2　Subjective Norms

Empirical evidence of the relationship between subjective norms and citizens' intention to use e-government services are found in many studies (Hung et al. 2006; Al Awadi and Morris 2009). Lau (2004) studed the adoption of e-government services in Hong Kong and found that the opinions held by family members have a significant influence on an individual's use of e-government services. Individuals within the society would influence others to use e- services to increase acceptance (Loch et al. 2003). Hung et al. (2006) found that both peers/friends/colleagues and the media have significant effects on subjective norms. McGrath et al. (2011) claim that online social networking in e-participation in Europe, the Middle East and Latin America allow the rapid mobilisation of citizens and the transfer of immediate information. Individual behaviour in a collectivist culture, such as that of Saudi Arabia, is affected by social norms received from people who are considered important to the individual. For-example, Al-Fulih (2002) shows that Saudi culture is socially active in the lives of the citizens and engenders strong relationships among family members. Therefore, family, friends/colleagues and media influence in this study are relevant referents of subjective norms and the intention to engage in e-participation. Therefore, the following hypotheses are developed.

**H5:** Family influence has a positive relationship with the subjective norm.

**H6:** Friends'/Colleagues' influences have a positive relationship with the subjective norm.

**H7:** Media influence has a positive relationship with the subjective norm.

**H8:** Subjective norm has a positive relationship with citizen's intention to engage in E-participation.

## 4　APPROACH

A quantitative research method was employed in this study and, specifically, a survey questionnaire was used for data collection. The survey (questionnaire) is one of the most common research methods in technology adoption studies as it uses a set of specific questions to cover the study topic and to target a large number of participants in a practical and efficient way (Carter and Bélanger 2005). For the validation and in order to test the hypotheses, data were collected from Saudi Arabian citizens. As the purpose of this study is to test e-participation in Saudi Arabia, the sample for this study was Saudi citizens who have experience in using the Internet and e-government websites. E-government website is the main way in which governments deliver government services through the web.

This study used an online survey method. The survey link was posted on social networking websites. The selection of this medium allowed for the inclusion of a culturally-diverse range of citizens and enabled the study to reach a large number of participants. The proposed model in this study includes nine constructs and multiple items on a five-point Likert scale to measure each construct. Previously validated survey measures were used in order to ensure the items were reliable. The questionnaire was originally developed in English; however, it was necessary to translate it into Arabic. Appendix A shows the survey measurement items of the factors in the study model.

### 4.1 Data Analysis

The data were analysed using a structural equation modelling (SEM) statistical technique, such as partial least squares (PLS) path modelling using SmartPLS version 3 (Ringle et al. 2014), to estimate the relationships between the different factors of the research model. SEM tests theoretical models using hypothesis testing to understand the simultaneous modelling of relationships among various independent and dependent factors. SEM is highly recommended in behavioural science and IS research (Gefen et al. 2000). This study followed a two-step approach to analyse the data (Anderson and Gerbing 1988): first, the measurement model was assessed to test convergent and discriminant validity and second, the structural model was assessed to test the research hypotheses.

## 5 FINDINGS

A total of 1233 responses were collected. After removing incomplete responses and outliers, a total of 770 samples were used to test the proposed model. Table 1 shows the demographics of the participants in this study in terms of gender, age, education level and Internet and e-government experience. Table 1 shows that the majority of the respondents are male (75.7%) and 24.3% are female. This gender imbalance could be related to the gender distribution of the users of e-government websites in Saudi Arabia (Abu Nadi 2012). The largest age group (46%) is made up of those between 20 and 29 years old. Almost half of the respondents have a Bachelor's degree education level (48.1%). 86% of participants have more than 5 years of Internet experience, while 11.6% have 3–5 years. Finally, 39% of participants have 1–3 years of experience using e-government websites, while 37% have more than 3 years.

| | **Characteristics** | **Frequency** | **Percent (%)** |
|---|---|---|---|
| **Gender** | Male | 583 | 75.7 |
| | Female | 187 | 24.3 |
| **Age** | Less than 20 years | 54 | 7 |
| | 20-29 | 359 | 46.6 |
| | 30-39 | 284 | 36.9 |
| | 40-49 | 59 | 7.7 |
| | 50year and above | 14 | 1.8 |
| **Education Level** | High School | 158 | 20.5 |
| | College degree | 92 | 11.9 |
| | Bachelor degree | 370 | 48.1 |
| | Postgraduate degree | 150 | 19.5 |
| **Internet experience** | Less than 1 year | 7 | 0.9 |
| | 1-3 years | 12 | 1.6 |
| | 3-5 years | 89 | 11.6 |
| | More than 5 years | 662 | 86 |
| **Experience in Using E-government Website** | Less than 6 months | 81 | 10.5 |
| | 7 – 12 months | 98 | 12.7 |
| | 1 – 3 years | 300 | 39 |
| | More than 3 years | 291 | 37.8 |

*Table 1: Demographic profile of respondents*

### 5.1 Measurement Model Assessment

The measurement model was evaluated by examining internal consistency, convergent validity and discriminant validity. The convergent validity relied on factor loading; the composite reliability (CR) and the average variance extracted (AVE). Constructs have convergent validity when, items loadings are above 0.5, the composite reliability exceeds the cut-of 0.70 and the average variance extracted (AVE) is above 0.50 (Hair et al. 2006). Table 3 shows the factor loadings, the average variance extracted (AVE), the composite reliability (CR), and the Cronbach Alpha values. All Items loadings were at least 0.70 and are more strongly on their assigned factor rather than on the other factor. The square root of the average variance extracted (AVE) for the constructs were above 0.50. Therefore, the results support the convergent validity of the constructs (Hair et al. 2006). In addition, all Cronbach Alpha values were more than 0.7, revealing good reliability. To evaluate the discriminant validity, the square root of the average variance extracted (AVE) for each construct was compared with the inter-factor correlations between a pair of the constructs. If the AVE is higher than the squared inter-scale correlations of the pair, then it indicates a good discriminant validity (Gefen et al. 2000; Hair et al. 2006). As shown in Table 2, the square root of AVE for each construct (factor) is greater than the correlations with other constructs. Therefore, the results support the discriminant validity of the model. Table 2 shows the Cronbach's reliability, composite reliability and the AVE of all constructs.

| **Factor** | CR | AVE | CAlpha | MID | TGO | TIT | ST | TRUT | INT | FAM | FRE | SN |
|---|---|---|---|---|---|---|---|---|---|---|---|---|
| MID | 0.845 | 0.649 | 0.831 | **0.805** | | | | | | | | |
| TGO | 0.951 | 0.710 | 0.951 | 0.372 | **0.842** | | | | | | | |
| TIT | 0.874 | 0.635 | 0.872 | 0.367 | 0.385 | **0.797** | | | | | | |
| ST | 0.887 | 0.723 | 0.885 | 0.170 | 0.185 | 0.313 | **0.851** | | | | | |
| TRUT | 0.830 | 0.621 | 0.827 | 0.287 | 0.506 | 0.459 | 0.396 | **0.788** | | | | |
| INT | 0.778 | 0.523 | 0.770 | 0.242 | 0.201 | 0.256 | 0.240 | 0.510 | **0.685** | | | |
| FAM | 0.895 | 0.739 | 0.894 | 0.422 | 0.339 | 0.372 | 0.258 | 0.361 | 0.310 | **0.860** | | |
| FRE | 0.900 | 0.750 | 0.899 | 0.508 | 0.280 | 0.353 | 0.192 | 0.333 | 0.350 | 0.640 | **0.866** | |
| SN | 0.887 | 0.664 | 0.888 | 0.458 | 0.274 | 0.384 | 0.233 | 0.447 | 0.498 | 0.467 | 0.602 | **0.815** |

*Notes:* **AVE**: Average Variance Extracted, **CR**: Composite Reliability, **CAlpha**: Cronbachs, **TGO**: Trust in Government, **TIT**: Trust in Internet, **ST**: Social Trsut, **TRUT**: Trust in e-participation, **FAM**: Family Influence, **FRI**: Friends Influence, **MID**: Media Influence, **SN**: Subjective Norm, **INT**: Intention to engage in e-participation

*Table 2: Measurement model results*

### 5.2 Structural Model Testing

The significance of the paths was determined by t-statistical test calculated with the bootstrapping technique at a 5 percent significance level. The coefficients of the causal relationships between factors were, by investigating the significance of the path coefficients and the ($R^2$) variance of the dependent variable. The coefficients and their t-value on the structural model, and the coefficients of determination ($R^2$) for dependent factors are shown in Table 3 and Figure 2.

| **Hypothesis** | **Path** | **Mean** | **SD** | **t-statistics** | **p-value** | **Supported** |
|---|---|---|---|---|---|---|
| H1 | TGO ➔ TRUT | 0.42 | 0.10 | 13.13 | 0.00*** | Yes |
| H2 | TIT ➔ TRUT | 0.24 | 0.11 | 7.70 | 0.02** | Yes |
| H3 | ST ➔ TRUT | 0.21 | 0.07 | 6.62 | 0.00*** | Yes |
| H4 | TRUT ➔ INT | 0.27 | 0.11 | 6.75 | 0.00*** | Yes |
| H5 | FAM ➔ SN | 0.40 | 0.07 | 12.15 | 0.06* | Yes |
| H6 | FRI ➔ SN | 0.43 | 0.09 | 13.38 | 0.04** | Yes |
| H7 | MID ➔ SN | 0.33 | 0.13 | 10.17 | 0.03*** | Yes |
| H8 | SN ➔ INT | 0.31 | 0.15 | 8.35 | 0.00*** | Yes |

*Notes:* **TGO**: Trust in Government, **TIT**: Trust in Internet, **ST**: Social Trust, **TRUT**: Trust in e-participation, **FAM**: Family Influence, **FRI**: Friends Influence, **MID**: Media Influence, **SN**: Subjective Norm, **INT**: Intention to engage in e-participation

***significant at the 0.001 level   **significant at the 0.01 level  *significant at the 0.05 level

*Table 3: Structural Model results*

As shown in the table 3, the result of this study confirms all hypotheses from H1 to H7 at p<0.05. $R^2$=0.37 indicates 37 percent variance in citizen trust. For the subjective norms $R^2$=0.53 indicates 53 percent variance in in subjective norms. $R^2$=0.29 indicates 29 percent variance in citizen intention to engage in e-participation in Saudi context. This means that trust in e-participation and subjective norms have significant effects on intention to engage in e-participation in Saudi context.

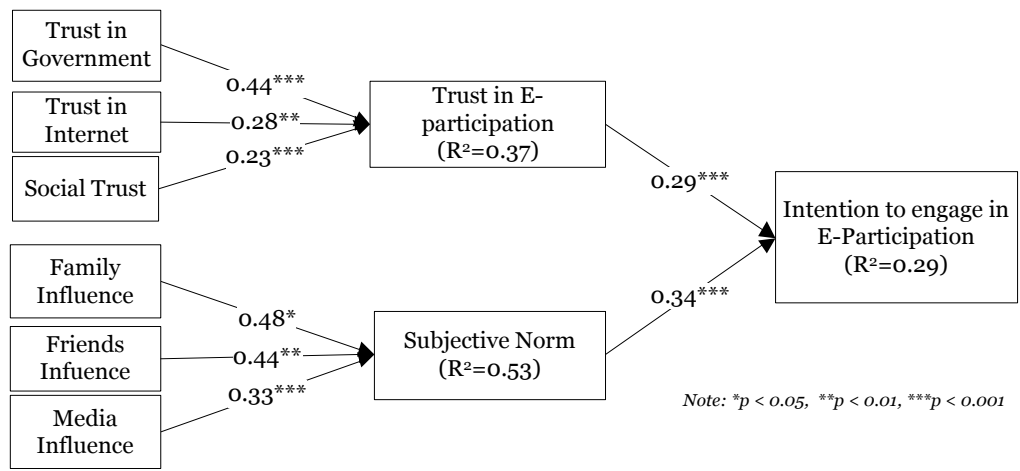

*Figure 2: Research Model Testing*

## 6  DISCUSSION AND CONCLUSION

The purpose of this research was to examine the influence of trust and subjective norms on citizens' intentions to engage in e-participation activities on e-government websites from the perspective of citizens in Saudi Arabia. According to the path testing (see Figure 2), trust in e-participation has a significant effect on citizens' intention to use e-participation. Further, the findings shows that trust in government and internet and social trust factors had a significant effects on citizens' trust in e-participation (in order of significance): 'Trust in the Government', 'Social Trust' and 'Trust in Internet'. Thus, 'Trust in Government' is the most important factor associated with 'trust in e-participation' in relation to citizens' 'intentions to engage in e-participation'. The subjective norm factors that had a significant effect on citizens' intentions to use e-participation (again, in order of significance) were: 'Media influence', 'Friends Influence' and 'Family Influence'. Thus, 'Media influence' is the most important factor associated with 'subjective norms' in relation to citizens' 'intentions to engage in e-participation'. The findings also showed that that 'subjective norms' had a greater effect on 'intention to use e-participation' than 'trust in e-participation'. Thus, citizen behaviours in the collectivist country of Saudi Arabia are more affected by the social norms of the people important to the participants.

The results also show that trust has a central role in creating effective conversations between citizens and governments for e-participation. In particular, trust in e-participation positively affected intended usage. These results are consistent with those of previous studies (Carter and Weerakkody 2008; Reddick 2011; Scherer and Wimmer 2014). In addition, citizens were found to be willing to engage in e-participation activities if their social norms are high towards e-participation because of other factors such as the opinions of friends and family members and media influence. This also is consistent with previous research (Gefen et al. 2006; AlAwadi and Morris 2009). The behaviour of citizens in Saudi Arabia was affected by the social influence of people considered important by the participants. Finally, the research model explains 39% of the variance of intention to engage in e-participation.

In addition, citizens are willing to engage in e-participation if their subjective norms are high towards e-participation because of other factors such the opinions of friends and family members and media influence. This is consistent with the research of ( Gefen et al. 2003; Hung et al. 2006; AlAwadi and Morris 2009). Saudi Arab is considered a high collectivistic culture (Hofstede et al. 2010) and therefore citizen behaviour in collectivistic country is affected by subjective norms received from people who are considered important to them. The results of this study show that citizens are willing to engage in e-participation activities if their subjective norms are high towards e-participation because of the opinions friends and family members or because e-government websites use a range of the tools for e-participation, such as social networking websites.

The finding shows that 'intention to engage in e-participation' is more affected by the subjective norms than the trust in e-participation. This might be attributable to the nature of e-participation. E-participation is optional for citizens, thus, the influence of others is higher in the Saudi context than 'trust in e-participation' in relation to citizens engaging in e-participation activities. Finally, the research model explains 29% of the variance of intention to engage in e-participation. This means people in Saudi are willing to use e-participation based on the trust and subjective norms determinants.

The key findings of this study will benefit the Saudi Arabian government (which has made substantial investments in e-government development initiatives to improve public services). In particular, the research highlighted the influence of key factors on how the Saudi citizens engage in e-participation activities on e-government websites. Thus, e-government officials should pay more attention to these factors to increase the level of e-participation use by its citizens. Trust is particularly relevant to the Saudi culture, as it is a culture characterized by a lack of trust and high power distance (Hofstede et al. 2010). The results showed that trust in government has an influence that is more significant on citizens' trust in e-participation than trust in the Internet and social trust. Thus, the Saudi government should focus on building confidence among citizens in the use of e-participation services. In relation to subjective norms, these findings are also important for the Saudi government and suggest that initiatives should be undertaken to encourage the citizens to engage in e-participation.

## 7 LIMITATIONS AND FUTURE WORK

Understanding the use of e-participation as a medium of communication between citizens and the Saudi government was the main purpose of this study. Moreover, the influence of trust and subjective norms on engaging in e-participation were considered particularly important in this context. The findings suggest that the significant determinants of trust (trust in the Internet, trust in government and social trust) and subjective norms (family, friends and media influence) should be the focus in the Saudi context. Due to the scope of this study, only the influence of trust and subjective norms on citizens' intention to engage in e-participation activities on e-government websites were investigated. Thus, the study did not provide a comprehensive examination of all the factors influencing the use of e-participation and future studies should therefore, consider other factors that may influence on citizens' intentions to engage in e-participation. With regard to the research sample, there was a gender imbalance in the participants who completed the survey. Future studies need to extend the sample size and employ other methods (such as focus group) to further validate the research model.

# Appendix

| Construct | Measurement Items | Sources |
|---|---|---|
| **Trust in Government** | The government agencies have the skills and expertise to provide e-participation activities in an expected manner. | Bhattacherjee (2002); Belanger and Carter (2005, 2008), Colesca (2009) |
| | The government agencies have the ability to meet the citizens' needs. | |
| | The government agencies can be trusted to participate in decisions | |
| | faithfully | |
| | The government agencies are truthful in its consulting with me via e-participation | |
| | I trust the government agencies to take my opinions in mind | |
| | I trust the government agencies to care about my opinions and suggestions | |
| | In my opinion, the government agencies are trustworthy | |
| **Trust in The Internet** | The internet has enough safeguards to make me feel comfortable to engage in e-participation activities on e-government websites | Belanger and Carter (2005, 2008) and Colesca (2009) |
| | I feel assured that legal and technological structures adequately protect me from problems on the Internet | |
| | I feel confident that encryption and other technological advances on the Internet make it safe for me to communicate with government agencies | |
| | In general, the Internet is a robust and safe environment to interact with government and other citizens | |
| **Social trust** | Most citizens are reliable | Pavlou and Gefen (2004); Colescav (2009) and Alsaghier et al. (2011) |
| | Most citizens keep commitments | |
| | Most citizens are honest in their opinions | |
| **Trust in E-participation** | I would trust e-participation on e-government websites to express my opinion right. | Colescav(2009) and Alsaghier et al. (2011) |
| | I trust e-participation on e-government websites. | |
| | I believe that e-participation in e-government websites is trustworthy. | |
| **Family Influence** | My family thinks that I should use e-participation in order to express my opinion to government | |
| | My family would think that using e-participation is a good idea | |
| | My family influenced me to try out e-participation | |
| **Friends/Colleagues Influence** | My friends/colleagues think that I should use e-participation on e-government websites in order to express my opinion. | |
| | My friends/colleagues would think that using e-participation is a good idea. | |
| | My friends/colleagues have an influence on me to try out e-participation. | Taylor and Todd (1995a); Bhattacherjee (2000), Hung, at al. (2006), Lin (2008) And Alzahrani (2011) |
| **Media Influence** | I read/saw news reports that using e-participation in e-government websites was a good way of expressing opinion/voice to government agencies | |
| | The popular press would depict a positive sentiment to engage in using e-participation in e-government websites | |
| | Mass media reports have an influence on me to try out e-participation | |
| **Subjective norms** | People who influence me think that I should use e-participation in e-government websites. | |
| | People important to me think that I should use e-participation in e-government websites. | |
| | People whose opinions I value would prefer that I use e-participation in e-government websites. | |
| | People who influence my decisions think that I should use e-participation in e-government websites. | |
| **Intention to engage in** | I intend to engage in e-participation activities on e-government websites | Carter and Bélanger (2005) |
| | I would engage in e-participation provided in e-government websites to participate in decision making. | |
| | Engaging in E-participation activities is something that I would do. | |
| | I would not hesitate to engage in e-participation activities on e-government websites to interact with government agencies. | |